\documentclass[11pt,aps,prb]{article}
\usepackage{color,graphicx,shortvrb,setspace}
  \usepackage[numbers,sort&compress,super,comma]{natbib}

\usepackage{float}
\usepackage{amssymb}
\usepackage{latexsym}

\newcommand{\be}{\begin{equation}}
\newcommand{\ee}{\end{equation}}
\newcommand{\bea}{\begin{eqnarray}}
\newcommand{\eea}{\end{eqnarray}}

\newcommand\fpk{\mbox{$f_{\mathrm{pk}}$}}
\def\ec{E_{\mathrm{c}}}

\def\nc{\nu_{\mathrm{c}}}
\def\ns{\mbox{$\nu^{\star}$}}
\def\ncs{\nu_{\mathrm{c}}^{\star}}

\newcommand{\rfig}[1]{Fig.\,\ref{#1}}
\newcommand{\rFig}[1]{Figure  \ref{#1}}
\newcommand{\rfigs}[2]{Figs.\,\ref{#1} and \ref{#2}}

\begin{document}
\title{Microwave spectroscopic observation of distinct electron solid phases in wide quantum wells
}
\maketitle
\author{\noindent  A.\,T. Hatke, Yang\,Liu, B.\,A. Magill, B.\,H. Moon, L.\,W. Engel, M. Shayegan, \\ L.\,N. Pfeiffer, K.\,W. West, K.\,W. Baldwin}

\vspace{.5in}
\noindent    National High Magnetic Field Laboratory, Tallahassee, Florida 32310, USA 

\vspace{.3in}
\noindent     Department of Electrical Engineering, Princeton University, Princeton, New Jersey 08544, USA

\newpage
\begin{doublespacing} 
 
{\bf In high magnetic fields ($B$), two dimensional electron systems (2DESs) can form a number of phases in which interelectron repulsion plays the  central role, since the kinetic energy is frozen out by Landau quantization. 
These phases include the well-known liquids of the fractional quantum Hall effect (FQHE), as well as solid phases with broken spatial symmetry and crystalline order. 
Solids can occur\cite{lozo,ee,hwj,vgwc,williams91,kunwc,jaincfwcnew,msreview} at the low Landau filling ($\nu$) termination of the FQHE series, but also within integer quantum Hall effects (IQHEs)\cite{chen:2003}.   
Here, we present microwave spectroscopy studies of wide quantum wells (WQWs) \cite{liu:2011,liu:2012,mssemi,haribi,suen:1994}.  
The spectra clearly reveal two distinct solid phases, hidden within what in dc transport would be the zero diagonal conductivity of an integer quantum Hall effect state.     
Explanation of these solids is not possible with the simple picture of a Wigner solid (WS) of ordinary (quasi) electrons or holes.
}

\begin{figure}[p]
\includegraphics[width=1\textwidth]{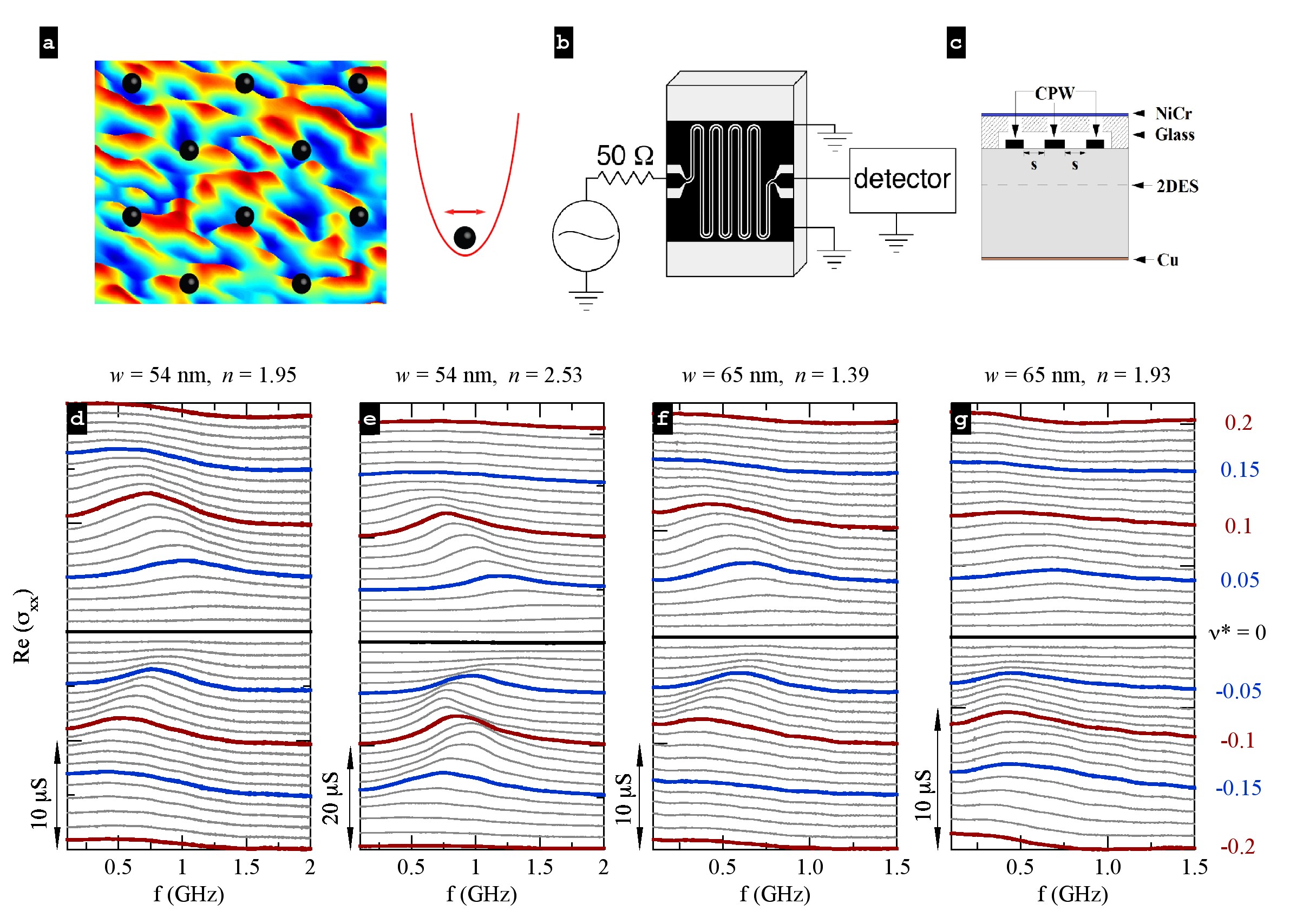}
\vspace{-0.2 in}
\caption{(a) Left: sketch of carriers (black) of an electron solid within the residual disorder potential which is shown by color (not to scale).     
Right: A piece (domain) of  solid moves within a potential due to the  disorder averaged over the domain, giving rise to the  pinning mode studied here. 
(b) Schematic of the microwave setup.   
The source and  detector are at room temperature.  
A coplanar waveguide (CPW) transmission line  is patterned onto the top surface of the sample, with metal film of the CPW shown in black.   
(c) Side view of the sample (not to scale) shows the placement of the front and back gates and  the metal film of the CPW.
Slots of width $s$ separate the center conductor and ground planes of the CPW.     
(d) and (e) Spectra (diagonal conductivity, Re$(\sigma_{xx})$, vs frequency, $f$) obtained from the 54 nm well with $n=1.95$ and $2.53$, at several $\ns=\nu-1$, marked on the right axis of (g).  
Successive spectra are offset upward by $1\mu$S in (d) and $2\mu$S in (e). 
(f) and (g) Spectra for the $65$ nm well at $n=1.39$ and $1.93$ offset consecutively by $1\mu$S.
}
\label{fcps}
\end{figure}

Solid phases of carriers are insulators, owing to pinning by  disorder, and are not easily   distinguishable from other types of insulators by standard dc transport measurements, but rather by a characteristic resonance, at frequency \fpk\ in their microwave conductivity spectra\cite{flr,williams91,chen:2003,chen:2004}. 
These resonances are understood as pinning modes\cite{flr,williams91} in which pieces of the solid oscillate within the disorder potential, as diagrammed in Fig.\,\ref{fcps}(a).
Pinning modes have been found both near the low $\nu$ termination of the FQHE series\cite{williams91,chen:2004} and within the $\nu$-ranges of IQHE plateaus\cite{chen:2003}. 
In the weak pinning picture\cite{fertig,chitra,foglerhuse}   \fpk\ increases when the shear modulus, $C_t$,  decreases, for example by decreasing the density, $ n $\cite{clidensity}.   
The inverse relation of \fpk\ and $C_t$  is because carriers associate more closely with minima in the disorder potential\cite{fertig,chitra,foglerhuse}.  
Consistent with this picture, in wells narrower than those studied here\cite{chen:2003}, a solid within the IQHE range shows  a monotonic decrease of \fpk\ as $\nu$ moves away from the quantizing filling, which increases the charge density of the solid.

Here we report pinning modes whose \fpk\ exhibits an {\em upward step} as $\nu$ is decreased from 1, in contrast to the monotonic decrease of \fpk\ that was seen in narrower wells.
The phenomenon is seen in WQWs, in which the effective electron-electron interaction is softened at short range due to the large growth-direction extent of the wave function.   
A wider well, or a larger $n$, cause the  step to move closer to $\nu=1$.       
We provide a natural interpretation, based on the sensitivity of \fpk\  to the properties of the solid, that the step signals a transition between different solids.

Our samples came from two WQW wafers with different width, $w$.
Henceforth in this paper all $n$ will be  in units of $10^{11}\,$cm$^{-2}$ for brevity. 
One WQW has $w=54\,$nm, depth of the 2DES from the top $d=430\,$nm, and as-cooled $n=2.42$.
The other has $w=65\,$nm, $d=510\,$nm, and as-cooled $n=1.52 $.  
The data were taken with a charge distribution that is approximately symmetric about the well center (see Methods). Microwave data were taken using the  set-up in Figs.\,\ref{fcps}(b) and (c) .

The spectra in Figs.\,\ref{fcps}(d)-(g)  give an overview of the evolution of the resonance with Landau filling at  different $n$ for the two WQWs.     
We present results in terms of $\ns=\nu-1$, where  $|\ns|$ is nearly  proportional to the quasiparticle or hole density.   
In the smaller-$n$ states in Figs.\,\ref{fcps}(d) and (f), and also for the larger-$n$ states in Figs.\,\ref{fcps}(e) and (g) with $\ns>0$, resonances develop as in Ref.\,\citenum{chen:2003}: as $|\ns|$ increases the resonance forms, develops maximal absorption around $|\ns|=0.08$ to $0.10$,  and then fades away, all as \fpk\ {\em monotonically} decreases with $|\ns|$.    
The resonance development is different for the larger-$n$ states with $\ns<0$.  
In Figs.\,\ref{fcps}(e) and (g) \fpk\ decreases with decreasing $\ns$ but near $\ns\sim -0.08$ begins to {\em increase}.
Further decrease of $\ns$ results in an increase of \fpk\ until $\ns\sim -0.12$, below which \fpk\ again decreases.

\begin{figure}[t]
\includegraphics[width=1\textwidth] {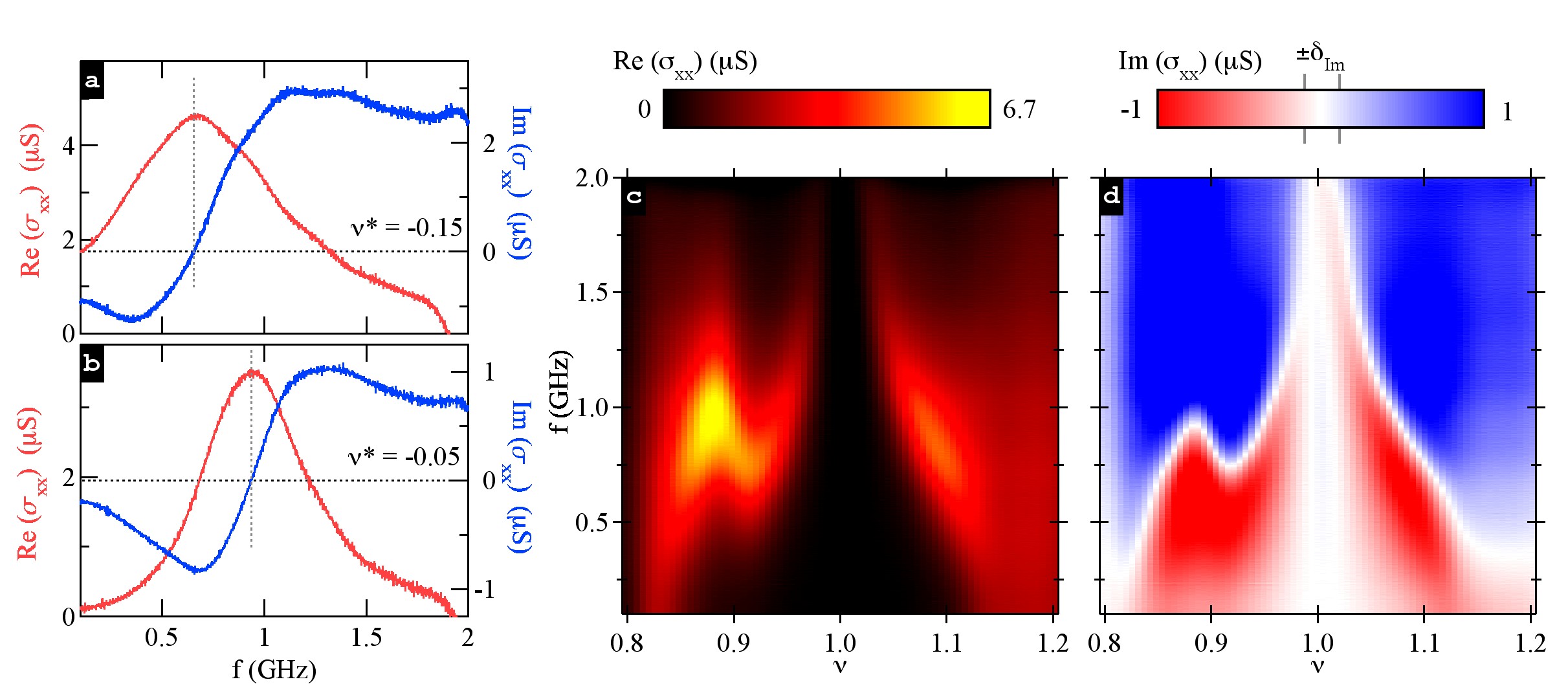}
\vspace{-0.2 in}
\caption{(a) and (b) Real  (left axis), and imaginary (right axis), parts of conductivity $\sigma_{xx}$ vs $f$ at two fixed $\nu^{*}$ as marked, obtained from the $54$ nm well at $n=2.53$. 
Dotted vertical lines mark the zero crossing of Im$(\sigma_{xx})$ vs $f$, from which \fpk\ is determined.
The error in Im$(\sigma_{xx})$, $\pm\delta_{\mathrm{Im}}=\pm 0.08\ \mu$S,   includes random noise and slower variation in Im$(\sigma_{xx})$ vs $f$ due to residual reflections near the sample mount.
(c) and (d) Color images in the $(f,\nu)$-plane for Re$(\sigma_{xx})$ and Im$(\sigma_{xx})$, respectively.
The color scale in (d) is set such that the center of the white band is 0, and its extent in Im$(\sigma_{xx})$ is $\pm\delta_{\mathrm{Im}}$.  
Thus the white band in the image traces out $\fpk$ vs $\nu$ and its height is $\pm\delta(\fpk)$. 
}
\label{error}
\end{figure}
\begin{figure}[t]
\vspace{-0.2 in}
\includegraphics{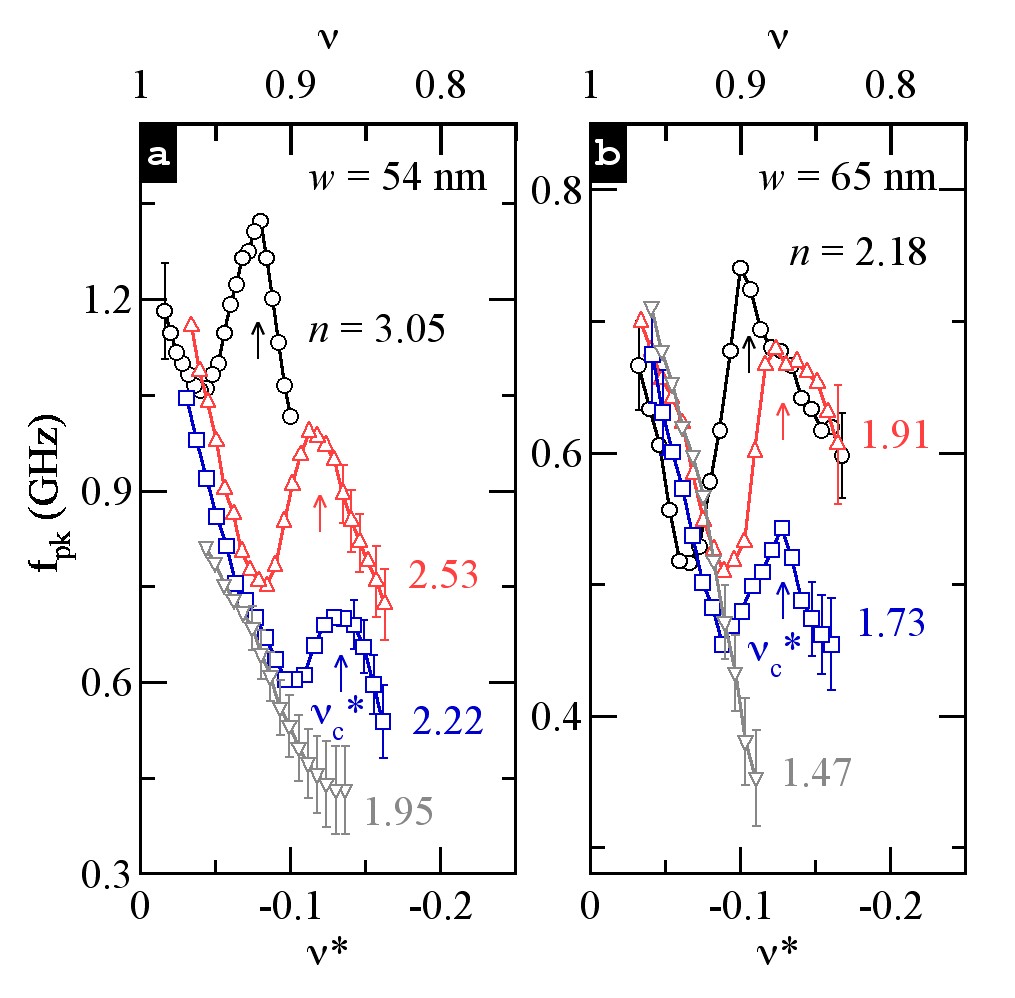}
 \vspace{-0.2 in}
\caption{$\fpk$ vs partial Landau level filling $\ns$ for the $54\,$nm (a) and $ 65\,$nm (b) wells at various carrier densities ($n$) as marked.   
Arrows mark $\ncs$, the \ns\ of the local maxima in \fpk. 
Data points without an attached error bar have estimated error in $\fpk$ that is smaller than  $\pm 5\%$.
}
\label{fpks} 
\end{figure}
$\fpk$ is extracted from the zero crossing of the imaginary conductivity, Im$(\sigma_{xx})$,  vs $f$, as explained in \rfig{error}.  
\rFig{fpks} shows plots of $\fpk$ vs $\ns$ at several $n$ for the two WQWs.  
For the lowest-$n$ traces, $\fpk$ vs $|\ns|$ monotonically decreases, but for larger $n$ a local minimum and maximum develop.  
The  minimum and maximum move closer to $\ns=0$ as $n$ increases and are exhibited at lower $n$ for the 65 nm sample.

We interpret the results in terms of a transition between two distinct  solids. 
At the lowest $n$ for both samples $\fpk$ monotonically decreases with $|\ns|$.   
At larger $n$, the \fpk\ vs $\ns$ curves can be divided into two regions. 
In the smaller $|\ns|$ regions (closer to $\nu=1$) \fpk\ vs $\ns$ tends toward the lowest-$n$ curve, while in the larger $|\ns |$ regions, to the right of the local maximum in \rfig{fpks}, \fpk\ vs $\ns$ is enhanced relative to the lowest-$n$ curve.       
We take the enhanced \fpk\ as a characteristic of a solid which we call S2.    It is distinct from S1, which is the only solid seen in the lowest $n$ states of the samples, and which in the larger $n$ states is closer to $\ns=0$.  
With increasing $n$ the transition from S2 to S1 moves closer to $\ns=0$, and the transition appears at lower $n$ in the larger $w$ sample.    
Screening of disorder by larger $|\ns|$, rather than a transition between solids, would produce a reduced \fpk\  for larger $|\ns|$, in contrast to the observed enhanced \fpk.
Though we focus on the resonance for $\ns<0$, we have, at some of the largest $n$, been able to see  the step to enhanced \fpk\ as $|\ns|$ increases for $\ns>0$ also.  
The transition thus appears to occur on both sides of $\nu=1$, therefore S2 is favored by larger $n$, $w$ and $|\ns|$.

\begin{figure}[p]
\includegraphics{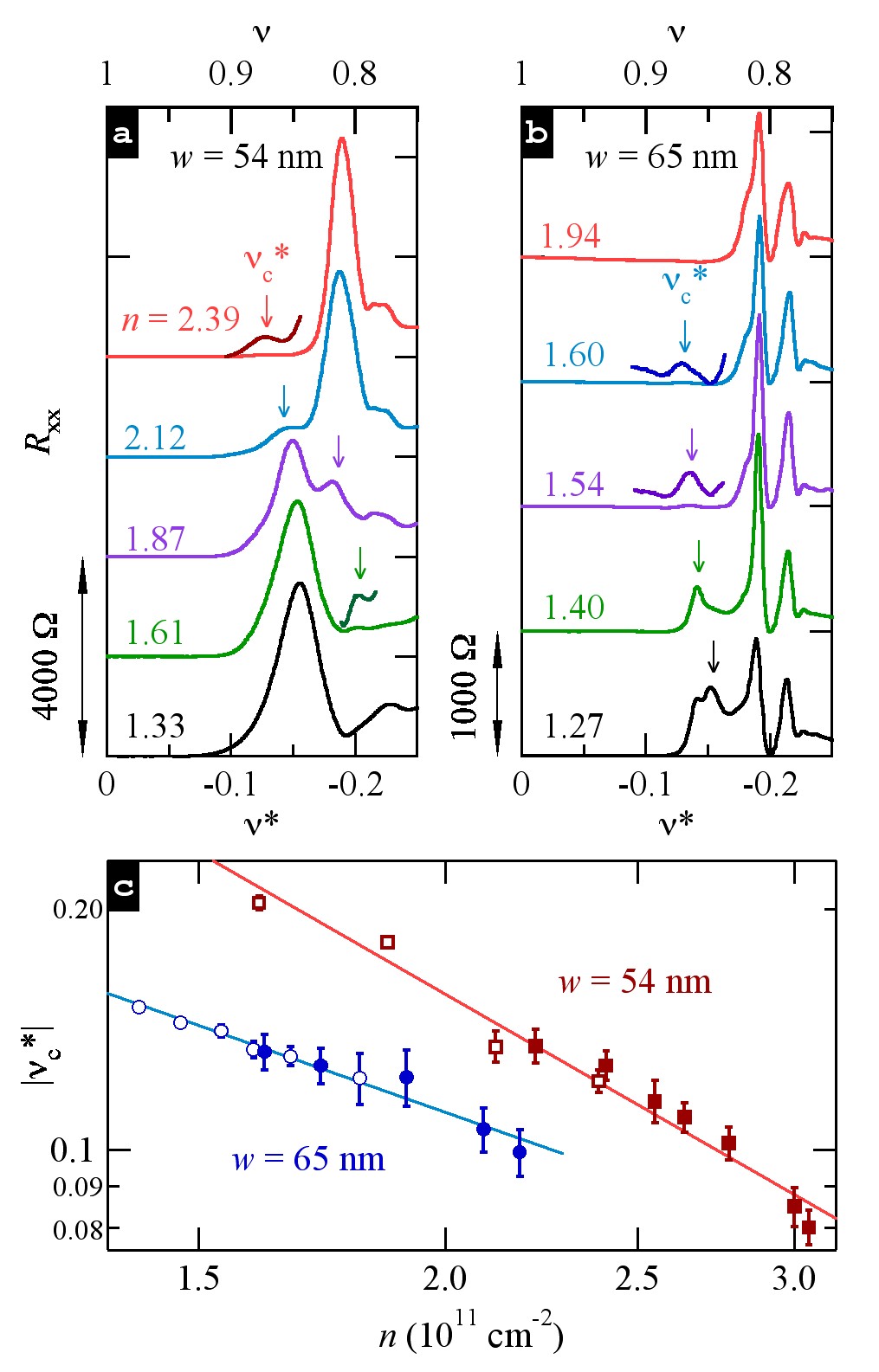}
\vspace{-0.2 in}
\caption{ 
(a) and (b): Longitudinal resistance versus $\ns$ measured for the $54\,$nm (a) and the $ 65\,$nm (b) wells at different $n$ as marked, vertically offset for clarity. 
Arrows mark $\ncs$, the critical partial filling factor.
Where $R_{xx}$ is small,  the trace in the region of $\ncs$ are duplicated and magnified 10 times  for clarity.
(c) $|\ncs|$ vs $n$ on a log-log scale deduced from dc transport measurements (open symbols) and microwave spectra (solid symbols) for the $54\,$nm (squares) and $65\,$nm (circles) wells.
Power law fits of the form $\ncs\propto n^{-\alpha}$ are also shown on the plot, where $\alpha=1.35\pm 0.1$ and $0.87\pm 0.07$ for the $54$ and $65\,$nm wells, respectively. 
}
\label{cn}
\end{figure}

We also perform dc transport measurements on the same wafers, and find that the solid-solid transition picture can underlie features found earlier\cite{liu:2012} in dc transport. 
WQWs were recently found\cite{liu:2012} to exhibit $\nu$ ranges that are not contiguous with the IQHE minimum centered at $\nu=1$, but still have Hall resistance quantized at $h/e^2$ and vanishing dc longitudinal resistance ($R_{xx}$).  
These regions are called reentrant integer quantum Hall effects (RIQHEs)\cite{jperiqhe} and are due to insulating phases of the partially filled Landau level.  
At sufficiently high $n$, the RIQHEs appear near $\nu=4/5$\,\cite{liu:2012}.
The RIQHE range extends toward $\nu=1$ as $n$ is increased by gating, and eventually merges with the main IQHE plateau.         
The RIQHE was ascribed\cite{liu:2012}  to a WS, which can be favored\cite{pricewqw} over FQHE liquids in WQWs, owing to the softening of the effective electron-electron interaction at short range.
The RIQHE sets in for $w/l_B\gtrsim 4$, where $l_B= (\hbar/eB)^{1/2}$ is the magnetic length.

Figures \ref{cn}\,(a) and (b) show plots of $R_{xx}$ vs $\ns$ for the $54$ and $ 65\,$nm wells at several $n$. 
Just closer to $\ns=0$ than the minimum due to the RIQHE, the $R_{xx}$  traces have peaks, both in Ref.\,\citenum{liu:2012} and in Figs.\,\ref{cn}\,(a) and (b), which are denoted with arrows.
With increasing $n$ the $\ns$ of the peak increases as the RIQHE minimum deepens.   
At the highest $n$ the peak vanishes as the RIQHE minimum merges with the main IQHE minimum.
This peak in $R_{xx}$ could be produced by domain-wall conduction at a phase transition\cite {liu:2012,gervaisdomain}.

To directly compare microwave and dc  measurements we define the critical filling $\ncs$ as that of the local maximum in the $\fpk$ vs $\ns$ measurements and that of the $n$-dependent peak in  $R_{xx}$ vs $\ns$, marked by arrows in \rfigs{fpks}{cn}.
This results in continuous plots of $\ncs$ vs $n$, as shown in \rfig{cn}\,(c), so that S2 can be identified with the dc RIQHE.   
The $\ns$-ranges of $\ncs$ vs $n$ obtained from microwave and dc overlap at intermediate $n$ and are remarkably consistent over the entire $n$ range. 
Resonances for S2 set in at larger $n$ than required to see the dc transport RIQHE, and  can be  seen at smaller $|\ns|$.   
When the RIQHE merges with the IQHE the $R_{xx}$ peak at $\ncs$ is no longer  visible, even when the transition is clearly visible in \fpk.   
Pinning modes are not visible in the RIQHE ranges at larger $|\ns|$, possibly because uncondensed carriers are present  which dampen the resonance.

Reference \citenum{chen:2004} reported evolution of  the pinning mode with $\nu$ below the termination of the FQHE series, in quantum wells with $w=50$ and $65\,$nm, but with much lower $n$ ($\simeq  1$ and $0.5 $ respectively) than those studied here.  For $0.12<\nu<0.18$,  the spectra appeared irregular with multiple peaks, though well-defined peaks appeared above and below that range.    
In Ref.\citenum{chen:2004} that $\nu$ range was the same in each sample, and for $\nu\gtrsim 0.18$,   the resonance was wave vector  ($q$) dependent.   
The $q$ dependence was found by comparing transmission lines of different slot width, $s$, $q \sim \pi/s$ (see Methods).   
Checking for $q$ dependence in the present $54\,$nm WQW sample  using $ s=80\ \mu$m  we found none, with excellent agreement between the $s=80\ \mu$m \fpk\ vs $\ns$ and the $s=30\ \mu$m data shown in \rfig{fpks}(a).

Theories\cite{jaincfwcnew,chang:2005,narevich:2001,yi:1998} have considered crystals of composite fermions (CFs).        
CFs are categorized by the number, $2p$, of vortices  bound to a carrier.   
Theory predicts a series of distinct CF WS ($^{2p}$CFWS) phases with $2p$ increasing as $\nu$ decreases\cite{jaincfwcnew,chang:2005}.
The  transition from $2p=2$ to $4$ occurs as $\nu$ goes from $1/5$ to  $1/6$.   
Though $^4C_t$ is predicted to soften near the transition, well within the $^4$CFWS it is calculated to be a factor of $\sim 2$ larger than $^2C_t$, and would result in a lower \fpk\ in the $^4$CFWS.    Identifying S2 as a $^2$CFWS and S1 as a $^4$CFWS would be consistent with the shear moduli   predicted by Ref.\citenum{jaincfwcnew}.   
However, Ref.\citenum{jaincfwcnew} predicts little sensitivity of the phase diagram to $w$, but we do observe dependence of $\ncs$ on $w$ and $n$.

Are there other possible interpretations for the two solids? 
A two-component bilayer WS can exist under certain conditions in a WQW\cite{haribi,mssemi}, particularly if the subband gap, $\Delta$, is small enough relative to the Coulomb energy, $\ec$.  
Such a two-component WS could have increased disorder and \fpk\ due to charge being pushed closer to the interfaces\cite{fertig}, and would be favored by larger $|\ns|$, giving smaller carrier spacing, and by larger $w$. 
For the $54\,$nm well with typical  $\nc=-0.1$, $n=2.8 $, we find from simulations\cite{liu:2011,mssemi} that $\Delta\sim16\,$K , which according to Ref.\citenum{narasimhan:1995} is about three times too large for a two-component lattice to form.

A CF ground-state spin transition\citep{du:1995,yeh:1999} is also unlikely to explain the observed phenomena.
The CF spin-Landau levels only cross above the Fermi level\cite{yeh:1999}.  
Though skyrmion solid formation has been reported\cite{hansk} from pinning modes near $\nu=1$, at  the larger $\ncs$ studied here large quasihole density would suppress skyrmion effects.

In summary, near $\nu=1$ in WQWs, we have found a change in \fpk\ that is naturally interpreted as signature of a transition between two different solids.       
S2, which exists at larger $|\ns|$, larger $n$ and larger $w$, has enhanced \fpk\ relative to the other phase.   
While the origin of the transition remains unclear, the possibilities, particularly of a transition in CFWS vortex number, are of fundamental importance.

\noindent {\em Acknowledgements}\\  
We thank Jainendra Jain for illuminating discussions.  
The microwave spectroscopy work at NHMFL was supported  through DOE grant DE-FG02-05-ER46212 at NHMFL/FSU.   
The National High Magnetic Field Laboratory (NHMFL),  is supported by NSF Cooperative Agreement No. DMR-0654118, by the State of Florida, and by the DOE.   
The work  at Princeton was funded through the NSF (grants DMR-1305691 and MRSEC DMR-0819860), the Keck Foundation and the Gordon and Betty Moore Foundation (grant GBMF2719).

\noindent {\em Methods}\\  
For dc measurements the symmetry of the  charge distribution about the well center was maintained using the Fourier transform of $R_{xx}$ vs $B$ in the Shubnikov-de Haas regime to minimize the gap, $\Delta$, between the lowest and first excited subbands\citep{suen:1994,liu:2011}.
For the microwave measurements, balance of the change between front and back halves of the well was maintained by biasing front and back gates such that individually each would change the carrier density by equal amounts.

Our microwave spectroscopy\cite{ee} technique \cite{chen:2003,chen:2004,hansk} uses a coplanar waveguide (CPW) on the surface of a sample.
A NiCr front gate was deposited on glass that was etched to space it from the CPW by $\sim 10\,\mu$m.
A schematic diagram of the microwave measurement technique is shown in the left panel of Fig.\,\ref{fcps}(b)  and  side view of the sample is shown  in Fig.\,\ref{fcps}(c).
From the power $P$ transmitted through the line, we calculate the diagonal conductivity as $ \sigma_{xx} (f) = (s/ l Z_{0}) \ln (t/t_0)$, where $s=30\ \mu$m is the distance between the center conductor and ground plane, $l=28\,$mm is the length of the CPW, and $Z_{0}=50\,\Omega$ is the characteristic impedance without the 2DES.  
$t_0$ is the  transmitted amplitude at $\nu=1$.   $\sigma_{xx} (f)$ is the difference between the  conductivity and that for $\nu=1$; just at $\nu=1$ the conductivity is vanishing at low temperature.  
The microwave measurements were carried out in the low-power limit, where  the measurement is not sensitive to the excitation power, and at a temperature of 60 mK.  
For dc transport measurements no transmission line was used; measurements were performed in a van der Pauw geometry using standard  ($\simeq 30\,$Hz) lock-in technique at a temperature of about 30 mK, with front and back gates both deposited directly on the sample surfaces. 

\noindent {\em Author Contributions}\\
A.T.H. performed the microwave measurements, analyzed the data and co-wrote the manuscript.
Y.L. designed, performed and analyzed the dc experiments and did numerical simulations.
B.A.M. and B.H.M. performed initial microwave measurements.
L.W.E. and M.S. conceived and designed the experiment, discussed data analysis, and co-wrote the manuscript.
L.N.P., K.W.W. and K.W.B. were responsible for the growth of the samples.

\end{doublespacing}

\bibliographystyle{nature}

\begin{thebibliography}{30}
\expandafter\ifx\csname natexlab\endcsname\relax\def\natexlab#1{#1}\fi
\expandafter\ifx\csname url\endcsname\relax
  \def\url#1{\texttt{#1}}\fi
\expandafter\ifx\csname urlprefix\endcsname\relax\def\urlprefix{URL }\fi

\bibitem[{Lozovik \& Yudson(1975)}]{lozo}
Lozovik, Y.~E. \& Yudson, V.
\newblock Crystallisation of a two dimensional electron gas in magnetic field.
\newblock \emph{JETP Letters} \textbf{22}, 11 (1975).

\bibitem[{Andrei(1988)}]{ee}
Andrei, E.~Y. \emph{et. al.}
\newblock Observation of a magnetically induced Wigner solid.
\newblock \emph{Phys. Rev. Lett.} \textbf{60}, 2765-2768 (1988).

\bibitem[{Goldman \emph{et~al.}(1990)Goldman, Santos, Shayegan \&
  Cunningham}]{vgwc}
Goldman, V.~J., Santos, M., Shayegan, M. \& Cunningham, J.~E.
\newblock Evidence for two-dimensional quantum Wigner crystal.
\newblock \emph{Phys. Rev. Lett.} \textbf{65}, 2189--2192 (1990).

\bibitem{hwj}H.-W. Jiang \emph{et~al.} 
\newblock Quantum liquid versus electron solid around $\nu=1/5$ Landau-level filling.
\newblock \emph{Phys. Rev. Lett}. \textbf{65}, 633 (1990). 

\bibitem[{Williams(1991)}]{williams91}
Williams, F.I.~B.  \emph{et. al.}
\newblock Conduction threshold and pinning frequency of magnetically induced Wigner solid.
\newblock \emph{Phys. Rev. Lett.} \textbf{66}, 3285-3288 (1991).

\bibitem[{Yang \emph{et~al.}(2001)Yang, Haldane \& Rezayi}]{kunwc}
Yang, K., Haldane, F. D.~M. \& Rezayi, E.~H.
\newblock Wigner crystals in the lowest Landau level at low-filling factors.
\newblock \emph{Phys. Rev. B} \textbf{64}, 081301 (2001).

\bibitem[{{Archer} \emph{et~al.}(2013){Archer}, {Park} \& {Jain}}]{jaincfwcnew}
{Archer}, A.~C., {Park}, K. \& {Jain}, J.~K.
\newblock {Nature of the crystal phase between 1/5 and 2/9 fractional Hall
  liquids}.
\newblock \emph{Phys. Rev. Lett.} \textbf{111}, 146804 (2013).

\bibitem[{Shayegan(1997)}]{msreview}
Shayegan, M.
\newblock \emph{{\rm in }Perspectives in quantum Hall effects, {\rm edited by
  S. Das Sarma and A. Pinczuk}}, 343 (Wiley-Interscience, New York, 1997).

\bibitem[{Chen \emph{et~al.}(2003)}]{chen:2003}
Chen, Y. \emph{et~al.}
\newblock Microwave resonance of the 2D Wigner crystal around integer Landau
  fillings.
\newblock \emph{Phys. Rev. Lett.} \textbf{91}, 016801 (2003).

\bibitem[{Liu \emph{et~al.}(2011)Liu, Shabani \& Shayegan}]{liu:2011}
Liu, Y., Shabani, J. \& Shayegan, M.
\newblock Stability of the $q/3$ fractional quantum Hall states.
\newblock \emph{Phys. Rev. B} \textbf{84}, 195303 (2011).

\bibitem[{Liu \emph{et~al.}(2012)}]{liu:2012}
Liu, Y. \emph{et~al.}
\newblock Observation of reentrant integer quantum Hall states in the lowest
  Landau level.
\newblock \emph{Phys. Rev. Lett.} \textbf{109}, 036801 (2012).

\bibitem[{Shayegan \emph{et~al.}(1996)Shayegan, Manoharan, Suen, Lay \&
  Santos}]{mssemi}
Shayegan, M., Manoharan, H.~C., Suen, Y.~W., Lay, T.~S. \& Santos, M.~B.
\newblock Correlated bilayer electron states.
\newblock \emph{Semiconductor Science and Technology} \textbf{11}, 1539 (1996).

\bibitem[{Manoharan \emph{et~al.}(1996)Manoharan, Suen, Santos \&
  Shayegan}]{haribi}
Manoharan, H.~C., Suen, Y.~W., Santos, M.~B. \& Shayegan, M.
\newblock Evidence for a bilayer quantum Wigner solid.
\newblock \emph{Phys. Rev. Lett.} \textbf{77}, 1813--1816 (1996).

\bibitem[{Suen \emph{et~al.}(1994)Suen, Manoharan, Ying, Santos \&
  Shayegan}]{suen:1994}
Suen, Y.~W., Manoharan, H.~C., Ying, X., Santos, M.~B. \& Shayegan, M.
\newblock Origin of the $\nu =1/2 $ fractional quantum Hall state in wide
  single quantum wells.
\newblock \emph{Phys. Rev. Lett.} \textbf{72}, 3405 (1994).

\bibitem[{Chen \emph{et~al.}(2004)}]{chen:2004}
Chen, Y.~P. \emph{et~al.}
\newblock Evidence for two different solid phases of two-dimensional electrons
  in high magnetic fields.
\newblock \emph{Phys. Rev. Lett.} \textbf{93}, 206805 (2004).

 \bibitem[{Fukuyama(1978)}]{flr}
 Fukuyama, H. \& Lee, P. A.
 \newblock Pinning and conductivity of two-dimensional charge-density waves in magnetic fields.
 \newblock \emph{Phys. Rev. B} \textbf{18}, 6245-6252 (1978).

\bibitem[{Fertig(1999)}]{fertig}
Fertig, H.~A.
\newblock Electromagnetic response of a pinned Wigner crystal.
\newblock \emph{Phys. Rev. B} \textbf{59}, 2120--2141 (1999).

\bibitem[{Chitra \emph{et~al.}(2001)Chitra, Giamarchi \& Le~Doussal}]{chitra}
Chitra, R., Giamarchi, T. \& Le~Doussal, P.
\newblock Pinned Wigner crystals.
\newblock \emph{Phys. Rev. B} \textbf{65}, 035312 (2001).

\bibitem[{Fogler \& Huse(2000)}]{foglerhuse}
Fogler, M.~M. \& Huse, D.~A.
\newblock Dynamical response of a pinned two-dimensional Wigner crystal.
\newblock \emph{Phys. Rev. B} \textbf{62}, 7553--7570 (2000).

\bibitem[{Li \emph{et~al.}(2000)}]{clidensity}
Li, C.-C. \emph{et~al.}
\newblock Microwave resonance and weak pinning in two-dimensional hole systems
  at high magnetic fields.
\newblock \emph{Phys. Rev. B} \textbf{61}, 10905--10909 (2000).

\bibitem[{Price \emph{et~al.}(1995)Price, Zhu, Das~Sarma \&
  Platzman}]{pricewqw}
Price, R., Zhu, X., Das~Sarma, S. \& Platzman, P.~M.
\newblock Laughlin-liquid\char21{}Wigner-solid transition at high density in
  wide quantum wells.
\newblock \emph{Phys. Rev. B} \textbf{51}, 2017--2020 (1995).

\bibitem[{Gervais \emph{et~al.}(2004)}]{gervaisdomain}
Gervais, G. \emph{et~al.}
\newblock Competition between a fractional quantum Hall liquid and bubble and
  Wigner crystal phases in the third Landau level.
\newblock \emph{Phys. Rev. Lett.} \textbf{93}, 266804 (2004).

\bibitem{jperiqhe}Lilly, M.~P.  \emph{et~al.}
\newblock Evidence for an anisotropic state of two-dimensional electrons in high Landau levels.
\newblock \emph{Phys. Rev. Lett.} \textbf{82}, 394 (1999).
\newblock The RIQHE was first seen in higher Landau levels, and is understood as a signature of bubble phases. 
Such phases are not predicted for the lowest Landau level, as discussed in Ref. 11.

\bibitem[{Chang \emph{et~al.}(2005)Chang, Jeon \& Jain}]{chang:2005}
Chang, C.-C., Jeon, G.~S. \& Jain, J.~K.
\newblock Microscopic verification of topological electron-vortex binding in
  the lowest Landau-level crystal state.
\newblock \emph{Phys. Rev. Lett.} \textbf{94}, 016809 (2005).

\bibitem[{Narevich \emph{et~al.}(2001)Narevich, Murthy \&
  Fertig}]{narevich:2001}
Narevich, F., Murthy, G. \& Fertig, H.~A.
\newblock Hamiltonian theory of the composite-fermion Wigner crystal.
\newblock \emph{Phys. Rev. B} \textbf{64}, 245326 (2001).

\bibitem[{Yi \& Fertig(1998)}]{yi:1998}
Yi, H. \& Fertig, H.~A.
\newblock Laughlin-Jastrow-correlated Wigner crystal in a strong magnetic
  field.
\newblock \emph{Phys. Rev. B} \textbf{58}, 4019 (1998).

\bibitem[{Narasimhan \& Ho(1995)}]{narasimhan:1995}
Narasimhan, S. \& Ho, T.-L.
\newblock Wigner-crystal phases in bilayer quantum Hall systems.
\newblock \emph{Phys. Rev. B} \textbf{52}, 12291 (1995).

\bibitem[{Du \emph{et~al.}(1995)}]{du:1995}
Du, R.~R. \emph{et~al.}
\newblock Fractional quantum Hall effect around $\nu = 3/2$: Composite fermions
  with a spin.
\newblock \emph{Phys. Rev. Lett.} \textbf{75}, 3926 (1995).

\bibitem[{Yeh \emph{et~al.}(1999)}]{yeh:1999}
Yeh, A.~S. \emph{et~al.}
\newblock Effective mass and $\mathit{g}$ factor of four-flux-quanta composite fermions.
\newblock \emph{Phys. Rev. Lett.} \textbf{82}, 592 (1999).

\bibitem[{Zhu \emph{et~al.}(2010)}]{hansk}
Zhu, H. \emph{et~al.}
\newblock Pinning-mode resonance of a skyrme crystal near Landau-Level filling
  Factor $\nu{}=1$.
\newblock \emph{Phys. Rev. Lett.} \textbf{104}, 226801 (2010).

\end{thebibliography}

\end{document}